\documentclass[11pt,a4paper]{article}
\usepackage{amsmath,amsfonts,amssymb,amscd}
\catcode`\@=11
\@addtoreset{equation}{section}

\catcode`\@=12

\newfont{\gotico}{eufm10 scaled\magstephalf}
\newfont{\qvd}{msam10 scaled\magstephalf}

\def\de#1/de#2{\frac{\partial {#1}}{\partial {#2}}}
\def\De#1/de#2{\dfrac{\partial {#1}}{\partial {#2}}}

\begin{document}
\vskip-2cm
\title{The most general ELKO Matter\\ in torsional $f(R)$-theories}
\author{Luca Fabbri$^{1,2}$\footnote{E-mail: luca.fabbri@bo.infn.it} \ 
and Stefano Vignolo$^{2}$\footnote{E-mail: vignolo@diptem.unige.it}\\
\footnotesize{$^{1}$INFN \& Dipartimento di Fisica, Universit\`{a} di Bologna}\\
\footnotesize{Via Irnerio 46, 40126 Bologna (Italia) and}\\
\footnotesize{$^{2}$DIPTEM Sez. Metodi e Modelli Matematici, Universit\`{a} di Genova}\\
\footnotesize{Piazzale Kennedy, Pad. D - 16129 Genova (Italia)}}
\date{}
\maketitle
\begin{abstract}
\noindent
We study $f(R)$-gravity with torsion in presence of the most general ELKO matter, checking the consistency of the conservation laws with the matter field equations; we discuss some mathematical features of the field equations in connection with a cosmological application.
\end{abstract}
\section{Introduction}
In the last decades, General Relativity has been extended toward several directions in order to solve some the problems left open by Einstein's theory in both the ultra-violet and the infra-red regime; among them one of the simplest is given by the so-called $f(R)$-theories: they consist in considering the gravitational Lagrangian to be a general function of the Ricci scalar $R$. This approach has acquired great interest in cosmology and astrophysics, where $f(R)$-theories turned out to be useful in addressing cosmological and astrophysical puzzles such as dark energy and dark matter: for example, they lead to possible explanations of the accelerated behaviour of the universe as well as the missing matter at galactic scales. General Relativity is also enlarged by considering torsion: that is the Ricci scalar $R$ is written in terms of the most general metric-compatible connection which carries torsional degrees of freedom. This geometry is enlarged enough to permit a corresponding generalization of physics, since having the background endowed with curvature and torsion allows the dynamical coupling to energy and spin: this is essential, because in the general theory of fields it is well-known that both energy and spin play an equally fundamental role.

The generalization of Einstein's theory obtained by introducing torsion was achieved by Cartan, and in the same way in which Einstein wrote the field equations coupling curvature to energy, Sciama and Kibble wrote the field equations coupling torsion to spin; the resulting theory known as Einstein-Cartan-Sciama-Kibble (ECSK) theory is variationally described by a gravitational Lagrangian linear in the Ricci scalar $R$ \cite{h-h-k-n,h-m-m-n}. Further generalization giving us an ECSK-like theory is the one for which the gravitational Lagrangian is non-linear in the Ricci scalar $R$: then, Einstein theory is in relationship with the ECSK theory in the same way in which the metric $f(R)$-theory is in relationship with the metric-torsional $f(R)$-theory \cite{CCSV1,CCSV2,CCSV3,CV}. Although in this last case torsion is present even without spin, nonetheless the matter fields that best exploit the coupling between torsion and spin density tensor are those having spin, that is the spinor fields; the simplest case is the spin-$\frac{1}{2}$ field, which in the case of Dirac fields it has been studied in \cite{f-v}.

However, recently a new form of spin-$\frac{1}{2}$ spinor field called ELKO has been defined; this form of matter gets its name from the acronym of the German \textit{Eigenspinoren des LadungsKonjugationsOperators} meaning ``eigenspinors of the charge conjugation operator'' defined as $\lambda$ for which $\gamma^{2}\lambda^{*}=\pm\lambda$ respectively for self- and antiself-conjugated fields \cite{a-g/1, a-g/2}: as a consequence of their definition they turned out to be fermions of mass dimension $1$ therefore described by scalar-like field equations \cite{a-l-s/1,a-l-s/2}. That ELKOs are fermions controlled by second-order derivative field equations is a fact that could lead to potential damages for the foundations of their dynamics; however the fundamental problems about acausality and singularities have been solved by showing that actually neither acausal propagation have place nor singularity formation occurs \cite{f/1, f/2, f/3}; as a consequence it makes sense to pursue the study of their dynamical properties by employing them in physical applications. In fact they too have gained a lot of interest in cosmology and astrophysics, where models with ELKOs were useful for the solution of cosmological and astrophysical issues such as dark matter and inflation: for instance, within these models there are promising explanations of the exponential expansion during the inflation of the universe and the constant velocity in the rotation curves of galaxies, as described in the literature reported here in references \cite{a} to \cite{g-m}. 

Quite recently, this theory of ELKOs has been generalized up to its most general dynamical structure in \cite{f}. In the present paper, we shall build the theory of $f(R)$ gravity with torsion coupled to fields of matter described by ELKOs in their most general form; our approach will face the problem of the consistency of matter field equations with the conservation laws, in the same way it has been done in \cite{f-v,f}. As it may be expected, the field equations of the theory will result to be rather complex, and in our discussion we shall point out some of the problems that could be faced: the main difficulty consists in the fact that the torsion-spin field equations will be in general differential equations for torsion, which is then a real dynamical variable, making the usual decomposition of the field equations in terms of Levi-Civita and torsional contributions hardly achievable, unless very special cases are considered; another important feature to highlight with respect to previous ELKO models is the fact that the presence of an additional dynamical term increases the intrinsic complexity of the theory, rendering its compatibility with too symmetric spacetimes difficult to accomplish. To exemplify these problems, in the end we shall consider a cosmological application to the case of a symmetric universe.

What we see to be one of the best advantages of having both ELKO and $f(R)$-theories is that it is possible to ascribe to two different sources the two different dark components of the universe, whose apparently opposite behaviour, attractive for dark matter and repulsive for dark energy, suggests that they are likely to have independent explanations. By encompassing these two theories into a single one may be fruitful for cosmology and astrophysics, but on the other hand, care must be taken in the choice of the underlying symmetries because, as we are going to show, highly symmetric spacetimes can be compatible with the flat solution alone.
\section{Geometrical Foundations}
In this paper, we shall indicate spacetime indices by Latin letters. A metric tensor on the spacetime is denoted by $g_{ij}$ and a connection by $\Gamma_{ij}^{\;\;\;h}$; metric-compatible connections are those whose covariant derivative applied on the metric tensor vanishes, where covariant derivatives are defined as
\begin{equation}
\label{2.1}
\nabla_{i}V_{j}=\partial_{i}V_{j}-\Gamma_{ij}^{\;\;\;h}V_{h}
\end{equation}
for any generic vector $V_{k}$. Given a connection $\Gamma_{ij}^{\;\;\;h}$, the associated torsion and Riemann curvature tensors are
\begin{subequations}
\label{2.2}
\begin{equation}
\label{2.2a}
T_{ij}^{\;\;\;h}
=\Gamma_{ij}^{\;\;\;h}-\Gamma_{ji}^{\;\;\;h},
\end{equation}
\begin{equation}
\label{2.2b}
R^{h}_{\;\;kij}
=\partial_i\Gamma_{jk}^{\;\;\;h} - \partial_j\Gamma_{ik}^{\;\;\;h} +
\Gamma_{ip}^{\;\;\;h}\Gamma_{jk}^{\;\;\;p}-\Gamma_{jp}^{\;\;\;h}\Gamma_{ik}^{\;\;\;p},
\end{equation}
\end{subequations}
where contractions $T_{i}= T_{ij}^{\;\;\;j}$, $R_{ij}=R^{h}_{\phantom{h}ihj}$ and $R=R_{ij}g^{ij}$ are called respectively the torsion vector, the Ricci tensor and the Ricci scalar curvature, and the commutator of covariant derivatives is expressed in terms of torsion and curvature as 
\begin{equation}
\label{2.3}
[\nabla_{i},\nabla_{j}]V_{k}=-T_{ij}^{\;\;\;h}\nabla_{h}V_{k}-R^{a}_{\;\;kij}V_{a}
\end{equation}
for any generic vector $V_{k}$. By considering the commutators of commutators in cyclic permutation and employing the Jacobi identities one obtains the Bianchi identities
\begin{subequations}
\label{2.4}
\begin{eqnarray}
\label{2.4b}
\nonumber
&\nabla_{c}T_{ij}^{\;\;\;h}-T_{ij}^{\;\;\;a}T_{ca}^{\;\;\;h}-R^{h}_{\;\;cij}
+\nabla_{i}T_{jc}^{\;\;\;h}-T_{jc}^{\;\;\;a}T_{ia}^{\;\;\;h}-R^{h}_{\;\;ijc}+\\
&+\nabla_{j}T_{ci}^{\;\;\;h}-T_{ci}^{\;\;\;a}T_{ja}^{\;\;\;h}-R^{h}_{\;\;jci}=0,
\end{eqnarray}
\begin{eqnarray}
\label{2.4a}
\nonumber
&\nabla_{c}R^{p}_{\;\;kij}-T_{ij}^{\;\;\;a}R^{p}_{\;\;kca}
+\nabla_{i}R^{p}_{\;\;kjc}-T_{jc}^{\;\;\;a}R^{p}_{\;\;kia}+\\
&+\nabla_{j}R^{p}_{\;\;kci}-T_{ci}^{\;\;\;a}R^{p}_{\;\;kja}=0.
\end{eqnarray}
\end{subequations}
Given a metric tensor $g_{ij}$ every metric $g$-compatible connection can be decomposed as
\begin{equation}
\label{2.5}
\Gamma_{ij}^{\;\;\;h}=\widetilde{\Gamma}_{ij}^{\;\;\;h}-K_{ij}^{\;\;\;h}
\end{equation}
so that
\begin{equation}
\label{2.6}
K_{ij}^{\;\;\;h}
=\frac{1}{2}\left(-T_{ij}^{\;\;\;h}+T_{j\;\;\;i}^{\;\;h}-T^{h}_{\;\;ij}\right),
\end{equation}
where $\widetilde{\Gamma}_{ij}^{\;\;\;h}$ is the symmetric Levi--Civita connection written in terms of the metric $g_{ij}$ alone and $K_{ij}^{\;\;\;h}$ is called the contorsion tensor, whose contraction $K_{i}^{\;\;ij}=K^{j}$ is such that $K_{i}=-T_{i}$; with the contorsion we can decompose the covariant derivative of the full connection as
\begin{equation}
\label{2.7}
\nabla_{i}V_{j}=\widetilde{\nabla}_{i}V_{j}+K_{ij}^{\;\;\;h}V_{h},
\end{equation}
where $\widetilde{\nabla}$ is the covariant derivative of the Levi--Civita connection and we can decompose the Riemann curvature of the full connection as
\begin{equation}
\label{2.8}
R^{k}_{\phantom{h}ihj}=\widetilde{R}^{k}_{\phantom{h}ihj}
+\widetilde{\nabla}_jK_{hi}^{\;\;\;k}-\widetilde{\nabla}_hK_{ji}^{\;\;\;k}+
K_{ji}^{\;\;\;p}K_{hp}^{\;\;\;k}-K_{hi}^{\;\;\;p}K_{jp}^{\;\;\;k}
\end{equation}
in terms of the Riemann curvature of the Levi--Civita connection $\widetilde{R}_{ij}$ identically.

In the next sections we shall consider spinor fields; as it is known, the most suitable variables to describe fermion fields are tetrad and spin-connections. Tetrad fields possess Lorentz indices denoted by Greek letters as well as spacetime indices denoted with Latin letters. They are defined by $e^{\mu}=e^{\mu}_{i}\,dx^{i}$ together with their dual $e_{\mu}=e_{\mu}^{i}\,\de /de{x^{i}}$, where $e^{j}_{\mu}e^{\mu}_{i}=\delta^{j}_{i}$ and $e^{j}_{\mu}e^{\nu}_{j}=\delta^{\nu}_{\mu}$ and such as they verify the orthonormality conditions $e^{i}_{\mu}e^{j}_{\nu}g_{ij}=\eta_{\mu\nu}$ and $e_{i}^{\mu}e_{j}^{\nu}g^{ij}=\eta^{\mu\nu}$ being $\eta$ the Minkowskian matrix of signature $(1,-1,-1,-1)$, while the spin-connection is defined as $1$-forms $\omega^\mu_{\;\;\nu} = \omega_{i\;\;\;\nu}^{\;\;\mu}\,dx^i$; we assume the metric compatibility conditions, the first given by the requirement that the covariant derivatives of tetrads vanish, yielding the relationships
\begin{equation}
\label{2.9}
\omega_{i\;\;\;\nu}^{\;\;\mu}
=\Gamma_{ij}^{\;\;\;h}e^\mu_h e_\nu^j-\partial_{i}e^{\mu}_{j}e_\nu^j
\end{equation}
that allows to employ the spacetime connection to calculate the spin-connection, while the second is given by the requirement that the covariant derivatives of Minkowskian matrix vanish, yielding the antisymmetry $\omega_{i}^{\;\;\mu\nu}=-\omega_{i}^{\;\;\nu\mu}$ of the spin-connection. In terms of the tetrads and the spin-connection the associated torsion and curvature tensors are written as
\begin{subequations}
\label{2.10}
\begin{equation}
\label{2.10a}
T^\mu_{ij}
=\partial_i\/e^\mu_j - \partial_j\/e^\mu_i
+\omega^{\;\;\mu}_{i\;\;\;\lambda}e^\lambda_{j}
-\omega^{\;\;\mu}_{j\;\;\;\lambda}e^\lambda_{i},
\end{equation}
\begin{equation}
\label{2.10b}
R_{ij}^{\;\;\;\;\mu\nu}
=\partial_i\omega_{j}^{\;\;\mu\nu} - \partial_j{\omega_{i}^{\;\;\mu\nu}}
+\omega^{\;\;\mu}_{i\;\;\;\lambda}\omega_{j}^{\;\;\lambda\nu}
-\omega^{\;\;\mu}_{j\;\;\;\lambda}\omega_{i}^{\;\;\lambda\nu},
\end{equation}
\end{subequations}
and they are related to the world tensors defined in equations \eqref{2.2} through the relationships given by $T_{ij}^{\;\;\;h}:=T^{\;\;\;\alpha}_{ij}e_\alpha^h$ and $R^{h}_{\;\;kij}=R_{ij\;\;\;\;\nu}^{\;\;\;\;\mu}e_\mu^h\/e^\nu_k$ respectively.

This concludes the introduction to the foundations of the geometry and the basic formalism, for which we also refer to \cite{h-h-k-n} for a more extensive discussion.
\section{Torsional $f(R)$-theories and conservation laws}
The torsional $f(R)$-theories can be formulated in the metric-affine approach \cite{CCSV1} or in the tetrad-affine one \cite{CCSV2}; in the first case, the gravitational dynamical fields are represented by the metric $g$ and a metric compatible connection $\Gamma$ while in the second case, the gravitational dynamical fields are given by a tetrad field $e^\mu_i$ and a spin-connection $\omega_{i}^{\;\;\mu\nu}$ on the spacetime. Field equations are derived variationally through a Lagrangian of the kind 
\begin{equation}
\label{3.1}
{\cal L}=f(R)\sqrt{|g|}dx_{1}\wedge dx_{2}\wedge dx_{3}\wedge dx_{4}-{\cal L}_m,
\end{equation}
where $f(R)$ is a real function of the Ricci curvature scalar $R$ written in terms of the metric and connection, or equivalently tetrad and spin-connection, and ${\cal L}_m$ indicates a suitable matter Lagrangian.

Because we are going to study the case in which there is a coupling with the spin density, namely we will study spinor fields, then the tetrad-affine formulation is more suitable: in this case the corresponding field equations are given by
\begin{subequations}
\label{3.3}
\begin{equation}
\label{3.3a}
f'(R)R_{\mu\sigma}^{\;\;\;\;\lambda\sigma}e^{i}_\lambda -\frac{1}{2}e^i_{\mu}f\/(R)=\Sigma^i_\mu,
\end{equation}
\begin{equation}
\label{3.3b}
f'(R)\left( T^\alpha_{ts} - T^\sigma_{t\sigma}e^\alpha_s + T^\sigma_{s\sigma}e^\alpha_t \right)
+\partial_{s}f'(R)e^\alpha_t -\partial_{t}f'(R)e^\alpha_s=S^\alpha_{ts},
\end{equation}
\end{subequations}
where $\Sigma^{i}_{\;\mu}:=-\frac{1}{2e}\frac{\partial{\cal L}_m}{\partial e^{\mu}_{i}}$ and $S_{ts}^{\;\;\;\alpha}:=\frac{1}{2e}\frac{\partial{\cal L}_m}{\partial\omega_{i}^{\;\;\mu\nu}}e^{\mu}_{t}e^{\nu}_{s}e^{\alpha}_{i}$ are the energy and spin density tensors of the matter field. From equation \eqref{3.3b} it is seen that, there are two sources of torsion given by the spin density $S_{ts}^{\;\;\;\alpha}$ and the non-linearity of the gravitational Lagrangian (for the derivation of the field equations and discussion about special properties of particular cases we refer to the works \cite{CCSV1,CV}).

It is then possible to write the field equations \eqref{3.3} in their equivalent spacetime form as 
\begin{subequations}
\label{3.5}
\begin{equation}
\label{3.5a}
f'(R)R_{ij}-\frac{1}{2}g_{ij}f(R)=\Sigma_{ij},
\end{equation}
\begin{equation}
\label{3.5b}
f'(R)\left(T_{ijh}+T_{j}g_{hi}-T_{i}g_{jh}\right)
+g_{hi}\partial_{j}f'(R)-g_{jh}\partial_{i}f'(R)=S_{ijh},
\end{equation}
\end{subequations}
where $R^i_{\;j}:=R_{\mu\sigma}^{\;\;\;\;\lambda\sigma}e^{i}_{\lambda}e^\mu_j$, $\Sigma^i_{\;j}:=\Sigma^i_{\mu}e^\mu_j$, $T_{ij}^{\;\;\;h}:=T^{\;\;\;\alpha}_{ij}e_\alpha^h\/$, $S^{\;\;\;h}_{ij}:=S^{\;\;\;\alpha}_{ij}e_\alpha^h\/$, which give the Ricci curvature tensor and torsion tensor in terms of the energy and spin densities. Notice that in equations \eqref{3.5a} one should distinguish the order of the indices since in general $R_{ij}$ and $\Sigma_{ij}$ are not symmetric.

Making use of the identities \eqref{2.4} it is possible to work out the field equations \eqref{3.5} to get the conservation laws of the theory
\begin{subequations}
\label{3.12}
\begin{equation}
\label{3.12a}
\nabla_{a}\Sigma^{ai}+T_{a}\Sigma^{ai}-\Sigma_{ca}T^{ica}-\frac{1}{2}S_{spq}R^{spqi}=0,
\end{equation}
\begin{equation}
\label{3.12b}
\nabla_{h}S^{ijh}+T_{h}S^{ijh}+\Sigma^{ij}-\Sigma^{ji}=0,
\end{equation}
\end{subequations}
which the energy and spin density tensors must fulfill once the matter field equations are assigned: the spin density is not conserved because of the antisymmetric part of the energy tensor whereas the energy is not conserved because of the presence of the spin-curvature coupling; therefore these have to be thought as conservation laws in a generalized sense.

For the derivation of the conservation laws in a variational context based on the Noether method see for instance \cite{h-m-m-n}, while for a direct derivation of the conservation laws explicitly based on the field equations see \cite{f-v}; in the next section we shall investigate in detail the coupling to the ELKO matter, showing that they indeed verify these conservation laws, as they should.
\section{Coupling to most general ELKOs\\ and consistency of field equations}
Let us consider $f(R)$-theories with torsion coupled to spinor fields, in the simplest spin-$\frac{1}{2}$ spin content; in \cite{f-v} we have already studied the Dirac field, here we would like to study the ELKO field. 

The ELKO matter fields $\lambda$ are defined as spin-$\frac{1}{2}$ spinor fields of Majorana type
\begin{eqnarray}
\gamma^{2}\lambda^{*}=\eta\lambda
\end{eqnarray}
with $\eta=\pm1$ for self- and antiself-conjugate fields; since the ELKO spinor can be decomposed in its two semi-spinors, then the $\lambda$ will be given in terms of one independent semi-spinor $L$ alone, and because of the definition above, the two components will turn out to have opposite helicities: so, although $\lambda$ cannot be an helicity eigenstate, nevertheless $L$ can be an helicity eigenstate, with either the positive of the negative eigenvalue. Indicating the two possible helicity eigenstates with $L_{\pm}$ we have that the ELKO may be denoted by
\begin{equation}
\begin{tabular}{c}
$\lambda_{\pm,\mp}=\left(\begin{array}{c}L_{\pm}\\ -\eta \sigma^{2}L_{\pm}^{*}
\end{array}\right)$
\end{tabular}
\end{equation}
identically. The ELKO dual is defined as 
\begin{equation}
\stackrel{\neg}{\lambda}_{\mp,\pm}:=\pm i\lambda_{\pm,\mp}^{\dagger}\gamma^{0}
\end{equation}
or explicitly
\begin{equation}
\stackrel{\neg}{\lambda}_{\mp,\pm}=\left(\begin{array}{cc}\mp i\eta L_{\pm}^{T}\sigma^{2} \ \ \ \ \pm i L_{\pm}^{\dagger}
\end{array}\right)
\end{equation}
as explained in \cite{a-g/1,a-g/2}; the definition of ELKO and ELKO dual endows them with additional properties under discrete transformations \cite{a-l-s/1,a-l-s/2}. All these properties for ELKO fields are important in making them Majorana spinors of peculiar type, and we refer to the papers above for a deeper understanding of the ELKO special algebraic structures. 

However, in the following we are not going to discuss these algebraic features any longer, because in this paper we are mainly interested in their differential character: since the ELKO matter field is a Majorana spinor, then the mass term would produce in their field equations all the well-known problems, usually avoided by the introduction of Grassmann variable, but which are solved in this case by the simple assumption of requiring for them second-order differential equations, which is what makes them special under a differential point of view.

Because ELKO fields have the same spin content of the Dirac field, that is they have the transformation law of any spin-$\frac{1}{2}$ particle, their spinorial covariant derivatives are defined in the same way by
\begin{eqnarray}
D_{i}\lambda=\partial_{i}\lambda+\omega_i^{\;\;\mu\nu}S_{\mu\nu}\lambda
\end{eqnarray}
with commutator of the derivatives given by
\begin{eqnarray}
[D_{i},D_{j}]\lambda=-T_{ij}^{\;\;\;h}D_{h}\lambda+R_{ij}^{\;\;\;\;\mu\nu}S_{\mu\nu}\lambda,
\end{eqnarray}
where $S_{\mu\nu}=\frac{1}{8}\left[\gamma_\mu,\gamma_\nu\right]$ and the gamma matrices $\gamma^{\mu}$ satisfy the anticommutation relationships given by the Clifford algebra, and we define $\gamma^i =\gamma^{\mu}e^i_\mu\/$ as usual.

Now from the fact that ELKO are spinors controlled by second-order field equations it respectively follows that they have a coupling to torsion and they have two derivatives in the field equations, so that their torsion tensor will be subject to derivation within the field equations, and back-reactions may therefore arise; consequently, problems in their causal propagation and issues for singularity formation have to be addressed \cite{f/1,f/2,f/3}, and additional terms may be used to generalize their dynamics as it has been shown in \cite{f}.

In this paper, we shall generalize their dynamics even further by generalizing their gravitational dynamics, and in this respect the most general Lagrangian for ELKOs in expressed as
\begin{equation}
\label{4.0}
{\cal L}_m=\left(D_{i}\stackrel{\neg}{\lambda}(g^{ij}+aS^{ij})D_{j}\lambda
-m^{2}\stackrel{\neg}{\lambda}\lambda\right)e \ dx_{1}\wedge dx_{2}\wedge dx_{3}\wedge dx_{4}
\end{equation}
in terms of the coefficient $a$ and where $m$ is the mass of the ELKO. 

By varying \eqref{4.0} with respect to the matter field we obtain the matter field equations
\begin{eqnarray}
\label{4.1}
\left(D^{2}\lambda+T^{i}D_{i}\lambda\right)
+a\left(S^{ij}D_{i}D_{j}\lambda+T_{k}S^{kj}D_{j}\lambda\right)+m^{2}\lambda=0
\end{eqnarray}
in terms of the mass $m$ of the matter field itself; by varying with respect to tetrads and spin-connection we get field equations \eqref{3.3} where the energy and spin density tensors are
\begin{subequations}
\label{4.2}
\begin{eqnarray}
\nonumber
\label{4.2a}
&\Sigma_{kj}=\frac{1}{2}\left(D_{j}\stackrel{\neg}{\lambda}D_{k}\lambda
+D_{k}\stackrel{\neg}{\lambda}D_{j}\lambda
-g_{jk}D_{i}\stackrel{\neg}{\lambda}D^{i}\lambda\right)+\\
&+\frac{a}{2}\left(D_{j}\stackrel{\neg}{\lambda}S_{ka}D^{a}\lambda
+D^{a}\stackrel{\neg}{\lambda}S_{ak}D_{j}\lambda
-g_{jk}D_{i}\stackrel{\neg}{\lambda}S^{ia}D_{a}\lambda\right)
+\frac{1}{2}g_{jk}m^{2}\stackrel{\neg}{\lambda}\lambda,
\end{eqnarray}
\begin{eqnarray}
\label{4.2b}
S_{kij}=\left(D_{j}\stackrel{\neg}{\lambda}S_{ki}\lambda
-\stackrel{\neg}{\lambda}S_{ki}D_{j}\lambda\right)
+a\left(D^{p}\stackrel{\neg}{\lambda}S_{pj}S_{ki}\lambda
-\stackrel{\neg}{\lambda}S_{ki}S_{jp}D^{p}\lambda\right),
\end{eqnarray}
\end{subequations}
and we will now explicitly verify that the matter field equations \eqref{4.1} are consistent with the conservation laws \eqref{3.12} in a direct way, without making use of the Lagrangian formalism \cite{h-m-m-n}.

To see this, we calculate the divergences of the conserved quantities
\begin{subequations}
\label{4.3}
\begin{eqnarray}
\nonumber
\label{4.3a}
&D_{k}\Sigma^{kj}=\frac{1}{2}(D^{k}D^{j}\stackrel{\neg}{\lambda}D_{k}\lambda
+D_{k}\stackrel{\neg}{\lambda}D^{k}D^{j}\lambda)+\\
\nonumber
&+\frac{a}{2}(D^{k}D^{j}\stackrel{\neg}{\lambda}S_{ka}D^{a}\lambda
+D^{a}\stackrel{\neg}{\lambda}S_{ak}D^{k}D^{j}\lambda)+\\
\nonumber
&+\frac{1}{2}
(D^{j}\stackrel{\neg}{\lambda}D^{2}\lambda+D^{2}\stackrel{\neg}{\lambda}D^{j}\lambda)+\\
\nonumber
&+\frac{a}{2}(D^{j}\stackrel{\neg}{\lambda}S^{ka}D_{k}D_{a}\lambda
+D_{k}D_{a}\stackrel{\neg}{\lambda}S^{ak}D^{j}\lambda)+\\
&+D^{j}(\frac{1}{2}m^{2}\stackrel{\neg}{\lambda}\lambda
-\frac{1}{2}D_{i}\stackrel{\neg}{\lambda}D^{i}\lambda
-\frac{a}{2}D_{i}\stackrel{\neg}{\lambda}S^{ia}D_{a}\lambda),
\end{eqnarray}
\begin{eqnarray}
\label{4.3b}
\nonumber
&D_{j}S^{kij}=(D^{2}\stackrel{\neg}{\lambda}S^{ki}\lambda
-\stackrel{\neg}{\lambda}S^{ki}D^{2}\lambda)+\\
\nonumber
&+a(D_{j}D_{p}\stackrel{\neg}{\lambda}S^{pj}S^{ki}\lambda
-\stackrel{\neg}{\lambda}S^{ki}S^{jp}D_{j}D_{p}\lambda)+\\
\nonumber
&+(D^{j}\stackrel{\neg}{\lambda}S^{ki}D_{j}\lambda
-D_{j}\stackrel{\neg}{\lambda}S^{ki}D^{j}\lambda)+\\
&+a(D_{p}\stackrel{\neg}{\lambda}S^{pj}S^{ki}D_{j}\lambda
-D_{j}\stackrel{\neg}{\lambda}S^{ki}S^{jp}D_{p}\lambda),
\end{eqnarray}
\end{subequations}
and by employing the matter field equations \eqref{4.1}, equations \eqref{4.3} simplify to
\begin{subequations}
\label{4.4}
\begin{eqnarray}
\nonumber
\label{4.4a}
&D_{k}\Sigma^{kj}=\frac{1}{2}(D^{k}D^{j}\stackrel{\neg}{\lambda}D_{k}\lambda
+D_{k}\stackrel{\neg}{\lambda}D^{k}D^{j}\lambda)+\\
\nonumber
&+\frac{a}{2}(D^{k}D^{j}\stackrel{\neg}{\lambda}S_{ka}D^{a}\lambda
+D^{a}\stackrel{\neg}{\lambda}S_{ak}D^{k}D^{j}\lambda)+\\
\nonumber
&+D^{j}(-\frac{1}{2}D_{i}\stackrel{\neg}{\lambda}D^{i}\lambda
-\frac{a}{2}D_{i}\stackrel{\neg}{\lambda}S^{ia}D_{a}\lambda)-\\
&-\frac{1}{2}T_{k}(D^{j}\stackrel{\neg}{\lambda}D^{k}\lambda
+D^{k}\stackrel{\neg}{\lambda}D^{j}\lambda)
-\frac{a}{2}T_{k}(D^{j}\stackrel{\neg}{\lambda}S^{ka}D_{a}\lambda
+D_{a}\stackrel{\neg}{\lambda}S^{ak}D^{j}\lambda),
\end{eqnarray}
\begin{eqnarray}
\label{4.4b}
\nonumber
&D_{j}S^{kij}=
a(D_{p}\stackrel{\neg}{\lambda}S^{pj}S^{ki}D_{j}\lambda
-D_{j}\stackrel{\neg}{\lambda}S^{ki}S^{jp}D_{p}\lambda)+\\
&+T_{j}(\stackrel{\neg}{\lambda}S^{ki}D^{j}\lambda
-D^{j}\stackrel{\neg}{\lambda}S^{ki}\lambda)
+aT_{j}(\stackrel{\neg}{\lambda}S^{ki}S^{jb}D_{b}\lambda
-D_{b}\stackrel{\neg}{\lambda}S^{bj}S^{ki}\lambda),
\end{eqnarray}
\end{subequations}
next we combine together the first three lines of equation \eqref{4.4a} and the first line of equation \eqref{4.4b} to get
\begin{subequations}
\label{4.5}
\begin{eqnarray}
\nonumber
\label{4.5a}
&D_{k}\Sigma^{kj}=\frac{1}{2}([D^{k},D^{j}]\stackrel{\neg}{\lambda}D_{k}\lambda
+D_{k}\stackrel{\neg}{\lambda}[D^{k},D^{j}]\lambda)+\\
\nonumber
&+\frac{a}{2}([D^{k},D^{j}]\stackrel{\neg}{\lambda}S_{ka}D^{a}\lambda
+D^{a}\stackrel{\neg}{\lambda}S_{ak}[D^{k},D^{j}]\lambda)-\\
&-\frac{1}{2}T_{k}(D^{j}\stackrel{\neg}{\lambda}D^{k}\lambda
+D^{k}\stackrel{\neg}{\lambda}D^{j}\lambda)
-\frac{a}{2}T_{k}(D^{j}\stackrel{\neg}{\lambda}S^{ka}D_{a}\lambda
+D_{a}\stackrel{\neg}{\lambda}S^{ak}D^{j}\lambda),
\end{eqnarray}
\begin{eqnarray}
\label{4.5b}
\nonumber
&D_{j}S^{kij}=
a(D_{p}\stackrel{\neg}{\lambda}[S^{pj},S^{ki}]D_{j}\lambda)+\\
&+T_{j}(\stackrel{\neg}{\lambda}S^{ki}D^{j}\lambda
-D^{j}\stackrel{\neg}{\lambda}S^{ki}\lambda)
+aT_{j}(\stackrel{\neg}{\lambda}S^{ki}S^{jb}D_{b}\lambda
-D_{b}\stackrel{\neg}{\lambda}S^{bj}S^{ki}\lambda).
\end{eqnarray}
\end{subequations}
By employing the commutators of spinorial covariant derivatives $D_{i}$ and the commutator of the generators $S_{ij}$ we obtain
\begin{subequations}
\label{4.6}
\begin{eqnarray}
\nonumber
\label{4.6a}
&D_{k}\Sigma^{kj}=-T^{kjh}\frac{1}{2}(D_{h}\stackrel{\neg}{\lambda}D_{k}\lambda
+D_{k}\stackrel{\neg}{\lambda}D_{h}\lambda)-\\
\nonumber
&-T^{kjh}\frac{a}{2}(D_{h}\stackrel{\neg}{\lambda}S_{ka}D^{a}\lambda
+D^{a}\stackrel{\neg}{\lambda}S_{ak}D_{h}\lambda)-\\
\nonumber
&-R^{abkj}\frac{1}{2}(\stackrel{\neg}{\lambda}S_{ab}D_{k}\lambda
-D_{k}\stackrel{\neg}{\lambda}S_{ab}\lambda)-\\
\nonumber
&-R^{abkj}\frac{a}{2}(\stackrel{\neg}{\lambda}S_{ab}S_{kp}D^{p}\lambda
-D^{p}\stackrel{\neg}{\lambda}S_{pk}S_{ab}\lambda)-\\
&-\frac{1}{2}T_{k}(D^{j}\stackrel{\neg}{\lambda}D^{k}\lambda
+D^{k}\stackrel{\neg}{\lambda}D^{j}\lambda)
-\frac{a}{2}T_{k}(D^{j}\stackrel{\neg}{\lambda}S^{ka}D_{a}\lambda
+D_{a}\stackrel{\neg}{\lambda}S^{ak}D^{j}\lambda),
\end{eqnarray}
\begin{eqnarray}
\label{4.6b}
\nonumber
&D_{j}S^{kij}=
\frac{a}{2}(D^{k}\stackrel{\neg}{\lambda}S^{ij}D_{j}\lambda
+D_{j}\stackrel{\neg}{\lambda}S^{ji}D^{k}\lambda
-D^{i}\stackrel{\neg}{\lambda}S^{kj}D_{j}\lambda
-D_{j}\stackrel{\neg}{\lambda}S^{jk}D^{i}\lambda)+\\
&+T_{j}(\stackrel{\neg}{\lambda}S^{ki}D^{j}\lambda
-D^{j}\stackrel{\neg}{\lambda}S^{ki}\lambda)
+aT_{j}(\stackrel{\neg}{\lambda}S^{ki}S^{jb}D_{b}\lambda
-D_{b}\stackrel{\neg}{\lambda}S^{bj}S^{ki}\lambda),
\end{eqnarray}
\end{subequations}
and according to the definition of the energy and spin density tensors \eqref{4.2} we finally obtain
\begin{subequations}
\label{4.7}
\begin{eqnarray}
\label{4.7a}
&D_{k}\Sigma^{kj}
=-T^{kjh}\Sigma_{kh}+\frac{1}{2}R^{abkj}S_{abk}-T_{k}\Sigma^{kj},
\end{eqnarray}
\begin{eqnarray}
\label{4.7b}
&D_{j}S^{kij}
=(\Sigma^{ik}-\Sigma^{ki})-T_{j}S^{kij},
\end{eqnarray}
\end{subequations}
showing that the matter field equations are consistent with the conservation laws. So the system of field equations given by the matter field equations \eqref{4.1} and the field equations \eqref{3.5} with conserved quantities \eqref{4.2} describes the most general ELKO matter in torsional $f(R)$-theories of gravity.

Now in dealing with equations \eqref{3.5} and \eqref{4.1}, the standard procedure consists in decomposing them in torsionless terms and torsional contributions, and more in detail the steps to follow are: firstly, obtaining from the trace of the Einstein-like equations \eqref{3.5a}, the expression of the Ricci scalar $R$ as a function of metric and matter fields with their derivatives; secondly, inserting the obtained relationship in the equations \eqref{3.5b} getting an explicit representation of the torsion tensor, again in terms of metric and matter fields with their derivatives; finally, replacing the expression for the torsion in equations \eqref{3.5a} by making use of equations \eqref{2.5}, \eqref{2.6} and \eqref{2.8}. By proceeding in this way, the theory can be reduced to an Einstein-like theory where the Einstein-like and matter field equations give the dynamics for the metric tensor and the matter fields; this procedure is useful in studying mathematical aspects such as the Cauchy, causality and singularity problems considered in \cite{CV1,CV2,CV3,f/1,f/2,f/3}.

In general for fermionic matter fields of the least-order derivative even within torsional $f(R)$ gravity this procedure works because the role of the torsion-spin equation \eqref{3.5b} is to define the torsion tensor as an algebraic function of the metric and matter fields together with their derivatives. However, this situation is rather different for ELKOs in their most general form in torsional $f(R)$ gravity, because both the energy and spin density tensors involve the covariant derivative of the spinors and when replacing the scalar curvature as function of the matter trace in the torsion-spin equation we no longer have an algebraic but a differential equation for torsion; in other words, we now have a dynamical equation for torsion which is then a genuinely dynamical variable.

Nevertheless, this procedure works for torsional $f(R)$ gravity in special cases such as $f(R)=R$; here in the case $a=0$ this decomposition is given explicitly in \cite{f/3}, while for the most general model $a\neq0$ the decomposition is not given explicitly although the fact that it is always possible to achieve is discussed through a constructive approach in \cite{f}. In general, the function $f(R)$ is not the identity, and the parameter $a$ does not vanish, but of course it could be possible to have special non-identical $f(R)$ functions and particular non-vanishing $a$ parameters, for which such a decomposition of torsion becomes achievable. Eventually, we will consider special symmetries, like those met in cosmology.

In the following we are going to choose the $f(R)$ function in a form that has already been considered in \cite{f-v} to be given by $f(R)=R-\varepsilon R^{2}$, while we choose the $a$ parameter to be given by $a=-2$; we consider the application to the spatially flat Friedmann-Lema\^{\i}tre-Robertson-Walker (FLRW) metric with line element of the following form 
\begin{equation}
ds^2=dt^2-e^{2\sigma}(dx^2+dy^2+dz^2)
\end{equation}
for $\sigma=\sigma(t)$: in this case, when the cosmological principle is implemented for torsion then of all irreducible decompositions only the two vectorial parts will not vanish, and for them, of all components only the temporal ones are different from zero, and we can choose to call them as
\begin{eqnarray}
T_{xyz}=T_{yzx}=T_{zxy}=6\varsigma,\\
T_{0}=3s,
\end{eqnarray}
in terms of $\varsigma=\varsigma(t)$ and $s=s(t)$. Finally, when the cosmological principle is implemented for matter, the ELKO fields can be written as the product of a real scalar function of the cosmological time and a constant ELKO, and therefore we may write them as
\begin{equation}
\label{ELKO}
\begin{tabular}{c}
$\lambda=\varphi\left(\begin{array}{c}0\\1\\i\\0\end{array}\right)$
\end{tabular}
\end{equation}
with dual
\begin{equation}
\label{ELKOdual}
\begin{tabular}{cccc}
$\stackrel{\neg}{\lambda}=\varphi\left(0 \ \ \ 1\ \ \ -i \ \ \ 0 \right)$,
\end{tabular}
\end{equation}
in terms of a $\varphi=\varphi(t)$ completely generic and where the constant ELKO has been chosen to have the form it has in order for the ELKO pseudo-scalar to vanish identically. In general, it is possible to build from ELKO and ELKO dual two invariants, being them the ELKO scalar and the ELKO pseudo-scalar, and therefore two special cases are found, when either the former or the latter vanish: for instance in \cite{b-b/1} the ELKO scalar is chosen to vanish, and thus ELKO ghosts are studied; instead in \cite{f} the ELKO pseudo-scalar is null, and hence ELKO fields are allowed to have energy although their energy tends to zero in the ultraviolet limit. So in this ultraviolet regime, the ELKO fields either tends to have zero energy, or they are ghosts never having energy, and thus the two conditions tend to coincide, giving rise to the same physics; but here no ultraviolet limit will be invoked, therefore the choice of the ELKO is not irrelevant, and because we prefer to study a situation in which ELKO fields are allowed to have energy in general, then the specific choice of the ELKO fields to be given in the form \eqref{ELKO} and \eqref{ELKOdual} above is imposed. The only choice that could have been even more general would have been in the case in which none of the two scalars were to vanish, but that more general case would have strongly increased the complexity of the treatment without really enriching the physical content. Our choice with ELKO pseudo-scalar null is physically meaningful and still simple enough to permit reasonable calculations. Now that the form of the ELKO, torsion and metric is chosen according to the arguments above and the cosmological principle, we may substitute them into the field equations. Field equations for the curvature are obtained by plugging \eqref{4.2a} into \eqref{3.5a} and they are equivalently written as
\begin{eqnarray}
\label{4.8}
\nonumber
&R_{jk}\left(1-2\varepsilon R\right)+\frac{1}{2}g_{jk}\varepsilon R^{2}
=\frac{1}{2}\left(D_{j}\stackrel{\neg}{\lambda}D_{k}\lambda
+D_{k}\stackrel{\neg}{\lambda}D_{j}\lambda\right)-\\
&-\left(D_{j}\stackrel{\neg}{\lambda}S_{ka}D^{a}\lambda
+D^{a}\stackrel{\neg}{\lambda}S_{ak}D_{j}\lambda\right)
-\frac{1}{2}g_{jk}m^{2}\lambda^{2}
\end{eqnarray}
after having substituted their contraction
\begin{eqnarray}
\label{4.9}
&R=D_{k}\stackrel{\neg}{\lambda}D^{k}\lambda
-2D_{k}\stackrel{\neg}{\lambda}S^{ka}D_{a}\lambda-2m^{2}\lambda^{2},
\end{eqnarray}
while the field equations for the torsion are obtained by plugging \eqref{4.2b} into \eqref{3.5b} and like torsion itself they can be decomposed into their two vectorial components and the only independent field equation is the trace
\begin{eqnarray}
\label{4.10}
&16s\left(1-2\varepsilon R\right)=\frac{d}{dt}\left(16\varepsilon R-\lambda^{2}\right)
\end{eqnarray}
in terms of the trace torsion since the axial field equation is satisfied by the fact that the axial torsion vanishes identically; finally we have that the field equations for the ELKO fields are not reduced any further by the choices we have made. Following the procedure we have outlined above, we plug \eqref{4.9} into \eqref{4.10} to get the expression of torsion that will have to be plugged back into all field equations to achieve the decomposition; however, even in this simplified case, calculations are very long although straightforward. To simplify the treatment, we notice that the spatial components of the spinorial derivatives are supposed to be proportional to the spatial momentum of the ELKO field which in cosmological models should vanish, and therefore we could require the vanishing of the spatial components of the spinorial derivative getting the condition $\dot{\sigma}=s$ as a restriction; this requirement however comes along with the vanishing of the curvature tensor $R^{h}_{\;\;kij}$: this circumstance implies that the form of the $f(R)$ is indistinguishable from the form $f(R)=R$ we would have in the simplest situation. In this simplest situation $f(R)=R$ the results have already been obtained in \cite{f}, in which it has already been proved that the ELKO field reduces to be constant, while the torsion and the purely metric curvature tensors vanish, and therefore this model is trivial.

It is now interesting to draw a parallel with precedent works about ELKO in cosmological applications, that is the one of Boehmer and Burnett \cite{b-b/1}, the previous \cite{f} and this one: in \cite{b-b/1} the authors consider the ELKO matter with energy and spin densities and show that they all are compatible with the cosmological principle; in \cite{f} however it has been shown that this most general model for ELKO gives rise to additional constraints that forces the theory to be more restricted by vanishing and flattening the geometrical background and leaving only an ELKO with no dynamical properties, and therefore the theory is trivial; here the most general ELKO in torsional $f(R)$ gravity gives rise to the flattening of the curvature tensor $R^{h}_{\;\;kij}$ forcing $f(R)$ to reduce to the simplest case, and thus recovering the trivial theory. This shows that the more we generalize the ELKO model or its underlying dynamical background the more we have problems of compatibility with too symmetric spacetime, forcing the matter fields to be more restricted unless the symmetries of the spacetime are loosened. Therefore if we still want to consider the ELKO matter in its utmost generality in general $f(R)$ gravity then the symmetries of the spacetimes must be weakened necessarily. Now the fact that the generalization we have proposed does not allow but an trivial model in the case of the FLRW spacetimes may appear to be unsatisfactory, but instead we think this should be regarded as a good property of our generalization. To explain why, we have to consider the physical interpretation that spinorial fields have for their spin-torsion coupling: intuitively, we may think that a spinor field, possessing spin, is endowed with rotational degrees of freedom; the fact that rotational degrees of freedom are coupled to the background suggests that the rotational degrees of freedom must be carried by the background. That a spinning field has its natural place in an axially symmetric background suggests that this type of matter should not give rise to backgrounds that are isotropic or isotropic and homogeneous, but on the other hand ELKO fields do, and without being trivial, as it has been shown in \cite{b-b/1}. This result is appealing because it constitute an unexpected, and therefore intriguing, circumstance, but on the other hand it also points toward the unexplained, and henceforth even more intriguing, situation for which this matter field should generate axial symmetries but it does not; we believe that the reason for this is that we have not been considering ELKOs in their most general case: in \cite{f} we have constructed the ELKO most general dynamics, and here we have placed that model in a background where the matter-geometry coupling given in terms of the Ricci scalar $R$ is the most general possible, the former generalization being represented by the parameter $a$ enriching the spin content, as explained in \cite{f}, the latter generalization being represented by the non-linearity of the function $f(R)$ enriching the spin-torsion coupling with an additional dynamical field equation, as explained above. And for ELKO in this most general situation it is not possible to have isotropic and homogeneous backgrounds, compatibly with the fact that matter fields should generate axial backgrounds if rotational degrees of freedom are to be express by their spin content adequately.
\section{Conclusions}
In this paper, we have considered $f(R)$-theories of gravitation with Ricci scalar written in terms of connections having both metric and torsional degrees of freedom, in the case in which the matter field was described by ELKOs in their most general dynamics: we have seen that the general conservation laws obtained in \cite{f-v} are satisfied for the energy and spin density tensors of ELKO once ELKO matter field equations are used; we have discussed general differences between ELKO and other matter fields in torsional $f(R)$-theories and between ELKO in this most general model and other simpler models previously studied.

It is known that both ELKO and $f(R)$-theories are very promising in explaining many of the open problems of cosmology and astrophysics, and ELKO fields in $f(R)$ gravity could give two different sources for the two complementary dark components of the universe; on the other hand, we have shown that these ELKO fields in their most general dynamics in $f(R)$ gravitation have an intrinsic complexity that makes them hardly compatible with too symmetric spacetimes: any cosmological model that wants to employ these generalized ELKOs must describe anisotropic universes at least in the initial epoch of their evolution. The study of this topic will be the subject of further works.

\end{document}